\documentstyle[prl,aps,epsf,rotate,multicol]{revtex}

\newcommand{\be}{\begin{equation}}
\newcommand{\ee}{\end{equation}}
\newcommand{\ba}{\begin{eqnarray}}
\newcommand{\ea}{\end{eqnarray}}

\begin{document}

\title{Connectivity Distribution of Spatial Networks}

\author{Carl Herrmann}
\address{
Dipartimento di Fisica Teorica dell'Universit\`a di Torino\\
and INFN sezione di Torino, 
Via P. Giuria 1, 10125 Torino, Italy\\
and\\
Dipartimento di Scienze e Tecnologie Avanzate
dell'Universit\`a del Piemonte Orientale\\
Corso Borsalino 54, I-15100 Alessandria, Italy.}

\author{Marc Barth\'elemy}

\address{
CEA, Service de Physique de la Mati\`ere Condens\'ee\\
BP12 Bruy\`eres-Le-Ch\^atel, France}

\author{Paolo Provero}

\address{
Fondazione per le Biotecnologie O.N.L.U.S.\\
Villa Gualino, Viale Settimio Severo 63, 10133 Torino, Italy
}

%\date{\today}
\maketitle
\begin{abstract}
We study spatial networks constructed by randomly placing nodes on a
manifold and joining two nodes with an edge whenever their distance is
less than a certain cutoff. We derive the general expression for the
connectivity distribution of such networks as a functional of the
distribution of the nodes. We show that for regular spatial densities,
the corresponding spatial network has a connectivity distribution
decreasing faster than an exponential. In contrast, we also show that
scale-free networks with a power law decreasing connectivity
distribution are obtained when a certain information measure of the
node distribution (integral of higher powers of the distribution)
diverges. We illustrate our results on a simple example for which we
present simulation results. Finally, we speculate on the role played
by the limiting case $P(k)\propto k^{-1}$ which appears empirically
to be relevant to spatial networks of biological origin such as the
ones constructed from gene expression data.
\end{abstract}

%02.50 -r stochastic processes
%05.40 -a fluctuation phenomena
%87.23.Ge   Dynamics of social systems 
%89.75.-k Complex systems
%89.75.Hc Networks and trees
%89.75.Fb Organization in complex systems
%87.23.Ge Dynamics of social systems

\begin{multicols}{2}

%%%%%%%%%%%%%%%%%%%%%%%%%%%%%%%%%%%%%%%

\section{Introduction}
In contrast to abstract graphs, many real networks are embedded in a
metric space: The interactions between the nodes depend on their
spatial distance and usually take place between nearest neighbors.
Examples of such networks are transportation and communication
networks, friendship or contact
networks\cite{Helmy:2002,Nemeth:2002}. An especially important example
is the Internet\cite{Lakhina:2002,Yook:2001}, which is a set of
routers linked by physical cables with different lengths.  Several
recent studies have investigated networks whose nodes are embedded in
a metric space, and where the probability of connecting two nodes with
an edge depends on their distance
\cite{Helmy:2002,Nemeth:2002,Yook:2001,Jost:2002,Barthelemy:2002,Dall:2002,Sen:2003,Warren:2002,ben-Avraham:2003}.

On the other hand, the concept of scale-free network has emerged in
the last few years as a powerful unifying paradigm in the study of
complex systems of natural, technological and social origin (see
Refs.~\cite{Albert:2001,Dorogovtsev:2001}). It is therefore natural to
investigate the possibility of embedding scale-free networks in space.
In particular in Refs.~\cite{Warren:2002,ben-Avraham:2003}, the
general problem of embedding a scale-free network of given
connectivity distribution in a Euclidean lattice was studied.

In this paper we take a somewhat reversed point of view: Our starting
point is a spatial distribution of points on a continuous manifold
$M$. Such points are the nodes of a network built by joining two nodes
whenever their distance is less than a certain cutoff. In the
following we will call `spatial networks' those obtained by this
procedure (sometimes also termed as `ad hoc networks'
\cite{Nemeth:2002}). We study how the connectivity distribution
depends on the distribution of the nodes on $M$. The fact that the
nodes live in a continuous manifold rather than a lattice is important
in generating scale-free networks since the number of neighbors that a
node can have within a certain distance is not limited {\it a priori}
by the lattice structure.

Spatial networks originating from a uniform distribution of the nodes
were studied in Ref.~\cite{Dall:2002}, where it was shown that while
the connectivity distribution takes the same Poisson form as in the
Erdos-Renyi random networks (see eg. \cite{Bollobas}), other important
features, most notably the clustering coefficient, are radically
different from the Erdos-Renyi case. The formation of giant clusters
in such networks was studied in Ref.~\cite{Dall:2002,Nemeth:2002}.

A natural application of this class of networks is the study of
epidemics propagation. While scale-free networks in which the
connectivity does not depend on a pre-existing metric structure do not
display an epidemic threshold \cite{Pastor-Satorras:2000}, the
situation changes when geographical closeness of two nodes influences
their probability to be connected \cite{Warren:2002}.  In our networks
geographical closeness {\it completely determines} whether two nodes
will be connected, so that we expect to find an epidemic
threshold. Spatial networks appear to be a rather realistic model for
epidemic propagation in animal or plant populations, where we do not
expect any individual to have an interaction range very different from
the average (as is the case for highly mobile human populations),
while we do expect the population density to be non-uniform.

In addition, spatial networks of biological origin have also recently
been constructed, especially from gene expression data obtained from
micro-array experiments \cite{Provero:2002,Herrmann:2002,Rho:2003}. The
connectivity distribution of such networks turns out to be a truncated
power-law with the exponent of the power-law decay often close to
unity. Interestingly, also networks constructed from gene expression
data by a different rule\cite{Agrawal:2002}, which does not satisfy
our definition of a spatial network, display a similar connectivity
distribution.

These facts motivated us to study how the connectivity distribution of
a spatial network depends on the distribution of the nodes in space
and under which conditions a scale-free network can be obtained, and
finally whether an exponent close to unity plays any special role in
this context.

%%%%%%%%%%%%%%%%%%%%%%%%%%%%%%%%%%%%%%%%%%%%%%%%%%
\section{The connectivity distribution of a spatial network}

\subsection{General Expression}

The $N$ nodes of the network are supposed to be in a $D$-dimensional
space and we will assume that they are distributed randomly in space
with density $p(x)$. Given a node chosen at random, the probability
that it is placed within a given domain of space $m$ is
\begin{equation}
\int_m dx\, p(x)
\end{equation}
(we denote the integration measure by $dx$ independently of the
dimension $D$).  Once the nodes are
distributed in this space, we have to construct the edges. We will use
a simple cut-off rule: Given two nodes $i$ and $j$, located at $x_i$
and $x_j$ respectively, an edge will join them if
\begin{equation}
d(x_i,x_j)\leq R\ \ ,
\end{equation}
where $d(x,y)$ is the distance between $x$ and $y$. Therefore, once
the nodes have been distributed the network is completely determined
by the choice of the cutoff $R$. This model follows strictly the rule
used to construct networks based on gene expression data in
Refs.~\cite{Provero:2002,Herrmann:2002,Rho:2003}, but more generally,
it can be used to model the case where the interaction has a typical
scale given by $R$.

Denoting by  $B_R(x)$ the ball of radius $R$ centered in $x$
\begin{equation}
B_R(x)\equiv \{y\in M: d(x,y)<R\}\ \ ,
\end{equation}
the probability that a given node is placed within
$B_R(x)$ is
\begin{equation}
q_R(x)=\int_{B_R(x)}dx^\prime\,p(x^\prime)\ \ .
\end{equation}

If we consider a node located at $x$, the probability that it will
have $k$ neighbors is then just the probability that $k$ additional 
nodes are
located in the $B_R(x)$. The connectivity distribution for a node
placed in $x$
is thus given by the binomial distribution
\begin{equation}
P(k;x,R)=\left(N-1\atop
k\right)q_R(x)^k\,\left[1-q_R(x)\right]^{N-1-k}\ .
\label{pkxr}
\end{equation}

In the following, we will be concerned with the limit $N\to\infty$:
In order to obtain a well defined connectivity distribution in this
limit, one has to take the limit $R\to 0$, so as to
ensure that the product
\begin{equation}
N\ V(R)
\end{equation}
where $V(R)$ is the volume of the ball
\begin{equation} 
V(R) = \int_{B_R(x)}dx\ \ ,
\end{equation}
tends to a finite constant $\alpha$. This means that $R$ must scale as
$N^{-D}$ as $N\to\infty$, and implies that the expected number of
nodes found within the ball $B_R(x)$ remains finite in this limit:
\begin{equation}
N\, q_R(x)\to \alpha\  p(x)\ \ .
\end{equation}
The constant $\alpha$ fixes the scale of the average connectivity
$\langle k\rangle$ of the network. Indeed, from Eq.(\ref{pka}) derived
below, it is easy to obtain the following relation between $\alpha$
and $\langle k\rangle$
\begin{equation} 
\langle k\rangle=\alpha\int dx\,p^2(x)
\end{equation}
Let us note that although the connectivity distribution is
well-defined for any value $\alpha$, it has been shown in the case of
a uniform density\cite{Dall:2002,Nemeth:2002} that the existence of a
giant connected component implies a minimum value of $\alpha$ (or
equivalently $\langle k\rangle$). We will not address this problem
here but we can expect that for other densities there will be some
similar conditions on $\alpha$ for the existence of a giant component.
 
In the limit $N\to\infty$, $R\to 0$, the connectivity distribution
for a node located at $x$ is Poissonian and Eq.(\ref{pkxr}) becomes
\begin{equation}
P(k;x,\alpha)=\frac{1}{k!}\alpha^k p^k(x) e^{-\alpha p(x)}\ \ .
\end{equation}
For the whole space, the connectivity distribution of the network is
then obtained as the spatial average of the former expression and is
\begin{equation}
P(k;\alpha)=\frac{\alpha^k}{k!}\int dx\, p^{k+1}(x) e^{-\alpha p(x)}
\label{pka}\ \ .
\end{equation}
This formula solves the general problem of determining the
connectivity distribution of a spatial network from the spatial
distribution of the nodes.  For a uniform distribution $p(x)$, we
recover a Poissonian connectivity distribution as in
Ref.~\cite{Dall:2002}. However, Eq.~(\ref{pka}) shows that other
connectivity distributions can be obtained. In the following section
we will determine under which condition a spatial node distribution
generates a scale-free network instead.

%%%%%%%%%%%%%%%%%%%%%%%
\subsection{Scale-free spatial networks}

Equation (\ref{pka}) allows us to determine the condition that must be
satisfied by the node density $p(x)$ for the corresponding network to
be scale-free. A network is scale-free if the moments $\langle
k^\nu\rangle$ of its connectivity distribution diverge for $\nu$
larger than a certain $\nu_{\rm max}$. It is easy to compute the
moments of the connectivity distribution of a spatial network from
Eq.~(\ref{pka}): For integer $\nu$ we have
\begin{eqnarray}
\nonumber\\
\langle k^\nu\rangle_\alpha &=& \sum_k \alpha^k \frac{k^\nu}{k!}
\int_M dx\, p^{k+1}(x) e^{-\alpha p(x)}
\nonumber\\
&=&\sum_{m=0}^\nu\sum_{k\ge m} \frac{\alpha^k}{k!}{\bf S}^{(m)}_\nu
(k)_m\int dx\,p^{k+1}(x)e^{-\alpha p(x)}\\
\nonumber\\
&=&\sum_{m=0}^\nu {\bf S}^{(m)}_\nu\alpha^m \int dx\,p^{m+1}(x)\ \ ,
\label{moments}
\ea 
where $(k)_m=k(k-1)\dots (k-m+1)$ and ${\bf S}^{(m)}_\nu$
are the Stirling numbers of the second kind (see
e.g. Ref.~\cite{Abramowitz}, p. 824), defined by 
\begin{equation}
x^\nu=\sum_{m=0}^\nu {\bf S}^{(m)}_\nu x(x-1)\dots (x-m+1)\ \ .
\end{equation}
It follows that $\langle k^\nu\rangle_\alpha$ exists if and only if
the integrals
\begin{equation}
H_a\equiv \int dx\,p^{a+1}(x)
\label{ha}
\end{equation}
exist for all $a\le \nu$. Conversely, the network is scale-free if
there exists a $\nu_{\rm max}$ such that $H_a$ diverges for all
$a\ge\nu_{\rm max}$. 
The integral Eq.~(\ref{ha}) is a measure of the
information contained in the probability distribution $p(x)$, and is
simply related to the Renyi entropy \cite{Renyi}
\begin{equation}
R_q = \frac{1}{1-q} \log H_{q-1}\ \ .
\end{equation} 

\subsection{Classes of spatial networks}
More generally, Eq.~(\ref{pka}) allows us to determine the type of spatial
network obtained for a given spatial node distribution.
The networks can essentially be distinguished by the decay of the
connectivity distribution for large connectivities\cite{Amaral:2000}. For
spatial networks, large connectivities are obtained in high density
regions, namely maxima of $p(x)$. For the sake of simplicity, we will
limit the discussion to the case of an isotropic distribution
$p(x)=p(r)$ where $r$ is the modulus of $x$. We will also suppose
that we have one density maximum located at
$r=0$. 
We will then distinguish the following two cases:
\begin{itemize}
\item{} $p(0)=p_0$ is finite and decays sufficiently rapidly so that
all the quantities $H_a$ are finite. This could be for instance the
case for population density which is decreasing exponentially from the
city center\cite{Clark:1951,Makse:1998}. In this case, an asymptotic
evaluation of the integral in Eq.~(\ref{pka}) shows that for large $k$
the connectivity distribution decays as
\begin{equation}
P(k;\alpha)\sim\frac{(\alpha p_0)^k}{k!k^D}\ \ .
\end{equation}
As expected, the low density fluctuations are reflected in the fast
decay of the connectivity distribution and the corresponding spatial
network will be of the `exponential' type\cite{Amaral:2000} (ie. the
connectivity distribution decreases at least as fast as an
exponential).

\item{} $p(r\sim 0)\sim r^{-\beta}$ with $\beta<D$ (if $\beta>D$ a
cut-off is needed to normalize $p(r)$ and we are in the first
situation where the maximum of $p(r)$ is finite). In this case, the
information measure $H_a$ will diverge for $a\geq\frac{D}{\beta}-1$
and the connectivity distribution is a power-law: Its large-$k$
behavior is given by (see section III)
\begin{equation}
P(k;\alpha)\sim k^{-D/\beta}\ \ .
\end{equation}
The large density fluctuations allow here for the existence of nodes
with very large connectivities and the corresponding network is
scale-free.
\end{itemize}
Even if it might appear unlikely that spatial densities behave
pathologically around some points, this is actually the case in many
instances where the nodes live in an abstract space, with a distance
defined by correlations.  Such examples are obtained in the case of
gene expression
networks\cite{Provero:2002,Herrmann:2002,Rho:2003,Agrawal:2002} (or
the stock market\cite{Mantegna:1998}) for which the distance is
defined in terms of Pearson correlation coefficient between nodes. The
spatial network has then a simple meaning: Two nodes are connected if
their correlation is high enough. It has been observed that in this
example, the connectivity distribution is a truncated power-law with
exponent of order $1$\cite{Provero:2002,Herrmann:2002,Rho:2003}.

\subsection{The dependence on the average connectivity}
In this section we consider the dependence of the connectivity
distribution (\ref{pka}) of a spatial network on the parameter
$\alpha$ that, as discussed in Sec.~2, is a measure of the average
connectivity. From Eq.(\ref{pka}) one immediately obtains the
following equation governing the dependence of $P(k;\alpha)$ on $k$:
\begin{equation}
\frac{dP(k;\alpha)}{d\alpha}=\frac1\alpha \left[ k\,P(k;\alpha)-
(k+1)\,P(k+1;\alpha)\right]\ \ .
\label{eeq}
\end{equation}
Note that while all spatial networks obey this equation, the converse
is not true: One can easily construct connectivity distributions
$P(k;\alpha)$ satisfying Eq.~(\ref{eeq}) that cannot be obtained from a
spatial network. 

Inspection of Eq.~(\ref{eeq}) shows that the limiting case $P(k)\propto
k^{-1}$ corresponds to a fixed point, where the connectivity
distribution does not depend on $\alpha$. 
This observation might be relevant in explaining the common appearance
of scale-free networks with connectivity exponent $\sim 1$ constructed
as spatial networks from gene expression data
\cite{Provero:2002,Herrmann:2002,Rho:2003}.

%%%%%%%%%%%%%%%%%%%%%%%%%%%%%%%%%%%%%%%
\section{A Simple Example of a Scale-free Spatial Network}

In this section we discuss an explicit example of a one-dimensional
distribution of nodes that generates a scale-free network. We
calculate exactly the connectivity distribution in the limit of
infinite number of nodes using Eq.~(\ref{pka}). We also address the
issue of finite-size effects and we compare the exact result for
infinite size to numerical simulation for finite networks.

\subsection{Connectivity distribution for an infinite network}

We consider the space to be the open interval $(0,1)$ and the
node distribution to be
\begin{equation}
p(x)=(1-\beta)\ x^{-\beta},\ \ \ \ \beta<1\ \ .
\label{node1d}
\end{equation}
The information measure $H_a$ is given by
\begin{equation}
H_a=(1-\beta)^{a+1}\int_0^1 dx\, x^{-\beta(a+1)}\ \ ,
\end{equation}
and diverges for 
\begin{equation}
a\ge\frac{1}{\beta}-1\ \ .
\end{equation}
Therefore we expect a scale-free network with $\langle
k^\nu\rangle_\alpha$ divergent for $\nu\ge\frac{1}{\beta}-1$. 

The connectivity distribution can be computed explicitly from
Eq.~(\ref{pka}) and is
\begin{eqnarray}
\nonumber\\
P(k;\alpha)&=&\frac{\alpha^k}{k!}(1-\beta)^{k+1}
\int_0^1dx\, x^{-\beta(k+1)}
e^{-\alpha(1-\beta)\ x^{-\beta}}\\
&=&\frac{1}{\alpha\beta}\left[\alpha(1-\beta)\right]^{1/\beta}
\frac{\Gamma\left[k+1-\frac1\beta,\alpha(1-\beta)\right]}{\Gamma(k+1)}\
\ ,
\label{pka1d}
\end{eqnarray}
where $\Gamma(x,a)$ is the incomplete Gamma function
\begin{equation}
\Gamma(x,a)=\int_a^\infty dt\, t^{x-1}e^{-t}\ \ .
\end{equation}
Since for $z\to\infty$
\begin{equation}
\frac{\Gamma(z,a)}{\Gamma(z)}\to 1\ \ ,
\end{equation}
we have for $k\to\infty$
\begin{eqnarray}
P(k;\alpha)
&\sim&
\frac{1}{\alpha\beta}\left[\alpha(1-\beta)\right]^{1/\beta}
\frac{\Gamma\left(k+1-\frac1\beta\right)}{\Gamma(k+1)} \label{exact}\\
&\sim&
\frac{1}{\alpha\beta}\left[\alpha(1-\beta)\right]^{1/\beta}
k^{-1/\beta}\ \ .
\label{scalefree}
\end{eqnarray}
This result shows explicitly that scale-free networks with any value
of the connectivity exponent down to 1 can indeed be obtained as a spatial
networks, as it is observed in gene expression networks.
Finally, we remark that the transition to a power-law behavior of the 
connectivity distribution (\ref{exact}) depends on $\alpha$ and $\beta$. 
Indeed, when the second argument $\alpha(1-\beta)$ 
of the incomplete gamma function tends to zero, we have a power-law behavior 
for $P(k;\alpha)$ even at small $k$, as can be seen from Fig.~(\ref{figure1}).

%%%%%%%%%%%%%%%%%%%%%%%%%%%%%%%%%%%
\subsection{Finite-size effects: Numerical results}

Since real spatial networks contain a finite number of nodes and our
analytical results were obtained in the limit of infinite networks, it
is important to address the issue of finite size effects. In this
section, we approach the problem from a numerical point of view by
constructing finite spatial networks by Monte Carlo methods and
comparing their connectivity distribution with the theoretical
predictions.

The figure $1$ shows the result of such comparison for the one-dimensional
spatial networks studied in the previous subsection, at
$\beta=0.5$ and $\beta=0.9$, and $\alpha=5$. For a finite network,
$\alpha$ is naturally defined as $2NR$,
where $N$ is the number of nodes and $R$ is the distance cutoff used
to define links. 
In each part of the figure, the connectivity distributions for
$N=1000$ and $N=20000$ points are superimposed to the theoretical
distribution given by Eq.~(\ref{pka1d}).
We can conclude that already for moderately sized networks the
connectivity distribution is very close to the $N\to\infty$ behavior
Eq.~(\ref{pka1d}), thus indicating that our results provide a good
approximation to the connectivity distribution of finite spatial networks.

\begin{figure}
\narrowtext 
\centerline{
\epsfysize=0.8\columnwidth{{\epsfbox{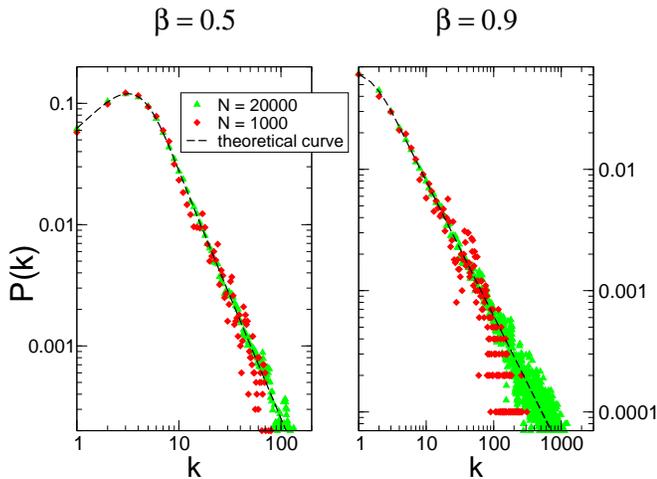}}}
}
\vspace*{.5cm}
\caption{ Monte Carlo simulation of the spatial networks
constructed from the one-dimensional node distribution (\ref{node1d})
compared to the $N\to\infty$ limit Eq.(\ref{pka1d}). The two figures
refer respectively to $\beta=0.5$ (0.9) with $\alpha=5$,  and show
for each case the result for $N=1000$ and $N=20000$.}
\label{figure1}
\end{figure}

%%%%%%%%%%%%%%%%%%%%%%%%%%%% discussion
\section{Discussion}

We have presented a systematic analysis of the connectivity
distribution of spatial networks constructed by joining nodes closer
to each other than a cutoff distance, in the limit in which the number
of nodes tends to infinity. The main results of our analysis can be
summarized as follows. First, the connectivity distribution can be
expressed as a functional of the spatial distribution of the nodes,
Eq.~(\ref{pka}). The moments of the connectivity distribution are
related to a certain measure of information $H_a$ of the node
distribution through Eq.~(\ref{moments}). In particular, scale-free
networks arise from this construction whenever the information measure
$H_a$ diverges for some $a$ and we showed that scale-free networks
with any exponent $\gamma>1$ can be constructed as spatial
networks. Our results were obtained in the limit of infinite number
$N$ of nodes but numerical results suggest that they provide an
excellent approximation of real, finite spatial networks already for
moderate $N$. Finally, the analysis of the dependence of the
connectivity distribution $P(k;\alpha)$ on the connectivity scale
$\alpha$ suggests that the limiting case $P(k)\propto k^{-1}$
corresponds to a fixed point, a fact that might be related to the
empirical observation that several spatial networks constructed from
gene expression data show a scale-free connectivity distribution with
exponent close to 1.

Intuition suggests that spatial networks will be {\it highly
clustered} and {\it highly assortative}. The clustering coefficient of
spatial networks was studied in Ref.~\cite{Dall:2002} for a uniform node
distribution: However since the clustering coefficient is a local
quantity and does not depend on the spatial density of nodes, the
results of \cite{Dall:2002} should hold unchanged for all spatial
networks, as long as the node distribution is isotropic in space.
Moreover, spatial networks can be expected to display a high degree of
assortativity, since high connectivity nodes are placed in high
density regions and are therefore more likely to be connected to other
high connectivity nodes. Indeed empirical spatial networks constructed
from gene expression data show a high degree assortativity
\cite{Herrmann:2002}. A closely related issue is the determination of
the diameter of spatial networks. Due to the embedding in a metric
space, we expect the diameter to grow as a power of the number of
nodes, and therefore we do not expect spatial networks to belong to
the small-world networks class.

Perhaps the most interesting open problem is to clarify
the role of the limit case $P(k)\propto k^{-1}$, namely to classify
the node distributions that flow to this fixed point in some limit. An
example is given by Eq.(\ref{pka1d}) in the $\beta\to 1$ limit.
$P(k)\propto k^{-1}$ implies that the measure of information $H_a$
diverges for all $a$, which in turn implies that the node distribution
must become degenerate (that is its support must shrink to zero
measure). However this is not a sufficient condition, since for
example a Gaussian $p(x)$ in the limit of zero variance, while
becoming degenerate, does not flow to $P(k)\propto k^{-1}$.

Acknowledgments: One of us (MB) thanks the department of physics-INFN
in Torino for its warm hospitality during the time this work was
completed.

%%%%%%%%%%%%%%%%%%%%%%%%%%%%%%%%%%%%%%%%%%%references

\end{multicols}

%%%%%%%%%%%%%%%%%%%%%%%%%%%%%%%%%%%%%%%figures

%%%%%%%%%%%%%%%%%%%%%%%%%%%%%%%%%%%%%%%%%%%%

%\end{multicols}
\end{document}